\documentclass[twocolumn,twoside,letterpaper,11pt]{article}
\usepackage{graphicx,xrrc,natbib}
\usepackage{times}
\def\percc{\mbox{$\;{\rm cm}^{-3}$}}
\def\gmpercc{\mbox{$\;{\rm gm}\;{\rm cm}^{-3}$}}

\def\mhz{\mbox{$\;$MHz}}     
\def\cms2{\mbox{$\;$cm s$^{-2}$}}    
\def\ergs{\mbox{$\;$ergs}}

\def\mug{\mbox{${\rm {\mu}G}$}}
\def\msun{\mbox{${{\rm M}_\odot}$}}

\def\arcsec{\rlap{$^{\prime\prime}$}\hbox to 2pt{}}
\def\arcmin{\rlap{$^{\prime}$}\hbox to 2pt{}}
\def\H0{\mbox{${\rm H_{0}}$}}
\def\mpc{\mbox{$\;{\rm Mpc}$}}

\def\kev{\mbox{$\;{\rm keV}$}}

\def\ev{\mbox{$\;{\rm eV}$}}

\def\yrs{\mbox{$\;{\rm yrs}$}}

\def\gtorder{\mathrel{\raise.3ex\hbox{$>$}\mkern-14mu
             \lower0.8ex\hbox{$\sim$}}}
\def\ltorder{\mathrel{\raise.3ex\hbox{$<$}\mkern-14mu
             \lower0.8ex\hbox{$\sim$}}}

\begin{document}


\title{Computer Simulations of Nonthermal Particles in Clusters of Galaxies:
Application to the Coma Cluster}


%
%
%
%


\author{Robert C.\ Berrington}
\institute{University of Wyoming}
\address{1000 E.\ University, P.\ O.\ Box 3905, Laramie, WY 82072}
\email{rberring@uwyo.edu}

\author{Charles D.\ Dermer}
\institute{Naval Research Laboratory} 
\address{Code 7653, Washington, DC 20375-5352} 
\email{dermer@ssd5.nrl.navy.mil}


\maketitle

\abstract{We have developed a numerical model for the temporal evolution of
particle and photon spectra resulting from nonthermal processes at the shock
fronts formed in merging clusters of galaxies. Fermi acceleration is
approximated by injecting power-law distributions of particles during a merger
event, subject to constraints on maximum particle energies. We consider
synchrotron, bremsstrahlung, Compton, and Coulomb processes for the electrons,
nuclear, photomeson, and Coulomb processes for the protons, and knock-on
electron production during the merging process. The broadband radio through
$\gamma$-ray emission radiated by nonthermal protons and primary and secondary
electrons is calculated both during and after the merger event.  To test the
ability of the computer model to accurately calculate the nonthermal emission
expected from a cluster merger event, we apply the model to the Coma cluster
of galaxies, and show that the centrally located radio emission and the Hard
X-ray excess observed at 40-80$\kev$ is well fit by our model.  If our model
is correct, then the Coma cluster will be significantly detected with GLAST
and ground-based air Cherenkov telescopes.  }
\section{Introduction}

Merger events between clusters of galaxies are extremely energetic events.
The formation history of a cluster of galaxies will include several merger
events \citep{gabici:03}.  With typical masses for a rich cluster
$\sim\!\!10^{15}\msun$, dynamical estimates of the energy deposited in to the
internal structure of a cluster is $\sim\!\!10^{63}$--$10^{64}\ergs$
\citep{sarazin:04}.  Approximately $5\%$ of the available energy is assumed to
accelerate particles from the thermal pool to create a distribution of
nonthermal particles.  The first-order Fermi process is capable of
accelerating particles to $\sim\!\!10^{19}\ev$ by the shocks that will form at
the interaction boundary of two merging cluster of galaxies
\citep{berrington:03}.

Optical and X-ray studies estimate that $\sim\!\!30$--$40\%$ of rich clusters
show signs of a current or recent merger event \citep{forman:81,beers:82}.
Structure formation calculations estimate that a cluster will see several
mergers though out its formation history \citep{lacey:93}, indicating that
most clusters contain a population of highly energetic nonthermal particles.

Diffuse cluster radio emission with no associated compact counterpart is
observed in an ever increasing number of galaxy clusters, providing evidence
for the prevalence of diffuse, highly energetic nonthermal particles in galaxy
clusters.  \citep{giovannini:99,kempner:01} The diffuse radio emission is
classified in to two categories.  Diffuse emission found in the cluster center
that mimics the thermal bremsstrahlung emission with random polarization is
known as a {\em radio halo}.  Diffuse cluster emission located on the cluster
periphery characterized by irregular shapes and strongly polarized light is
known as a {\em radio relic}.  These features are preferentially seen in
clusters with current or recent cluster merger events, and are thought to be
the observational consequences of nonthermal particles accelerated by shocks
formed in a cluster merger event.

\section{Models}

We present results of a numerical model \citep{berrington:03} designed to
calculate the time-dependent evolution of nonthermal particle distribution
functions evolving through radiative losses.  The electrons and are
accelerated by the first-order Fermi process and at the shock fronts formed in
merging clusters of galaxies, and the resulting photon radiation is
calculated.  Particle injection functions $Q_{e,p}$ for electrons (``e'') and
protons (``p'') are assumed to be described by power-law momentum spectra. In
terms of kinetic energy $K_{e,p}$, the injection functions are given by
$$Q_{e,p}(K_{e,p},t) = Q_{e,p}^{0} [K_{e,p} (K_{e,p} + 2 m_{e,p}
c^{2})]^{-\frac{s + 1}{2}}$$
\begin{equation}
(K_{e,p} + m_{e,p} c^2) \exp\left[ - \frac{K_{e,p}}{K_{\rm max}} \right],
\label{eqn:power-law}
\end{equation}
where $s$ is the injection index and $K_{\rm max}$ is the maximum particle
energy determined by three conditions: the available time to accelerate to a
given energy since the formation of the cluster merger event; the requirement
that the particle Larmor radius is smaller than the size scale of the system;
and the condition that the energy-gain rate through first-order Fermi
acceleration is larger than the energy-loss rate due to synchrotron and
Compton processes.  The constant $Q_{e,p}^{0}$ normalizes the injected
particle function, and is determined by
$$\int_{K_{\rm min}}^{K_{\rm max}} dK_{e,p} ~ K_{e,p}
Q_{e,p}(K_{e,p},t) = $$
\begin{equation}
\frac{\eta_{e,p}}{2} A \; \eta^{e}_{\rm He} m_{p} v_{1}^{3} \langle n_{\rm
ICM} \rangle\;,
\label{eqn:normalization}
\end{equation}
where $\langle n_{\rm ICM} \rangle$ is the number density of the
intra-cluster medium (ICM) averaged over the area A of the shock front,
$\eta^{e}_{\rm He}$ is an enhancement factor to account for the ions heavier
hydrogen, $\eta_{e,p}$ is an efficiency factor taken to be $5\%$ unless
otherwise noted, and $m_{p}$ is the mass of a proton.  The minimum kinetic
energy $K_{\rm min}$ is held constant at $10\kev$.  

The model calculates the forward [$v_{1}$] and reverse [$v_{2}$] shock speed
from the gravitational infall velocity $v_{g}$ of a merging cluster. The
trajectory of the smaller merging cluster is approximated by a point mass of
total mass $M_{2}$ that falls onto a dominant cluster of total mass $M_{1}$
whose density profile is described by an isothermal beta model.  The velocity
$v$ of the shock fluid is calculated by solving the equation
\begin{equation}
\frac{\mu_{1}}{\mu_{2}} \frac{n_{1}}{n_{2}} = \frac{1 + 3{\cal
    M}_{1}^{-2}}{1 + 3 {\cal M}_{2}^{-2}} \left( \frac{v_{g} - v}{v}
    \right)^{2}\;,
\label{eqn:relative_shock_velocity}
\end{equation}
where $n_{1}$ and $n_{2}$ are the number densities in the dominant and merging
cluster, respectively.  The mean atomic mass in the dominant cluster and the
merging cluster are given by $\mu_{1}$ and $\mu_{2}$, respectively.  Both mean
atomic masses are set equal to $0.6m_{p}$.

The Mach speeds of the forward [${\cal M}_{1}$] and reverse [${\cal
M}_{2}$] shocks are calculated by 
$${\cal M}_1 = \frac{2} {3} \frac{v}
{c_{1}} \left( 1 +
\sqrt{1 + \frac{9} {4}
\frac{c_{1}^{2}} {v^{2}} } \right) \;, $$
\noindent and
\begin{equation}
{\cal M}_2 = \frac{2} {3} \frac{v_{g} - v} {c_{2}} \left( 1 +
    \sqrt{1 + \frac{9} {4} \frac{c_{2}^{2}} {(v_{g} - v)^{2}} } \right)\;,
\label{eqn:forward_reverse_mach_number}
\end{equation}
where $c_{1}$ is the sound speed in the dominant cluster, and $c_{2}$ is the
sound speed in the merging cluster.  The Mach number is defined to be ${\cal
M}_{1,2} = v_{1,2}/c_{1,2}$ for the forward and reverse shock, respectively.
Equation \ref{eqn:relative_shock_velocity} and equation
\ref{eqn:forward_reverse_mach_number} are derived from the shock jump
conditions, by equating the energy densities of the forward- and
reverse-shocked fluids at the contact discontinuity.  Compression ratios
$C_{1}$ (forward) and $C_{2}$ (reverse) are calculated from the equation
\begin{equation} 
C_{1,2} = \frac{4}{ 3 (1-{\cal M}_{1,2}^{-2})}
\label{eqn:shock_velocity}
\end{equation}

The time-dependent particle spectrum $N(K,t)$ is determined from solving the
Fokker-Planck equation in energy space, given by
$$\frac{\partial N(K,t)}{\partial t} = \frac{1}{2} \frac{\partial^{2}}{\partial
K^{2}} \left[D(K,t)~N(K,t)\right]$$
$$- \frac{\partial}{\partial K}
\left[\dot{K}(K,t) N(K,t)\right] $$
\begin{equation} -
\sum_{i=pp,p\gamma,d}\frac{N(K,t)}{\tau_{i}(K,t)} + Q(K,t).
\label{eqn:fokker_planck}
\end{equation}
Here $\dot{K}(K,t)$ is the total kinetic-energy loss rate found by the sum of
Coulomb, synchrotron, bremsstrahlung and Compton processes for electrons, and
Coulomb processes for protons.  In addition, the protons experience
catastrophic losses from proton-proton ($i=pp$) collisions, proton-$\gamma$
($i=p\gamma$) collisions, and diffusive escape ($i=d$) on the timescale
$\tau_{i}(K,t)$.  Secondary electrons are calculated and added to the primary
electron distribution function and are subject to the same energy losses as
the primary electrons.

\begin{figure}
\begin{center}
\includegraphics[width=\columnwidth]{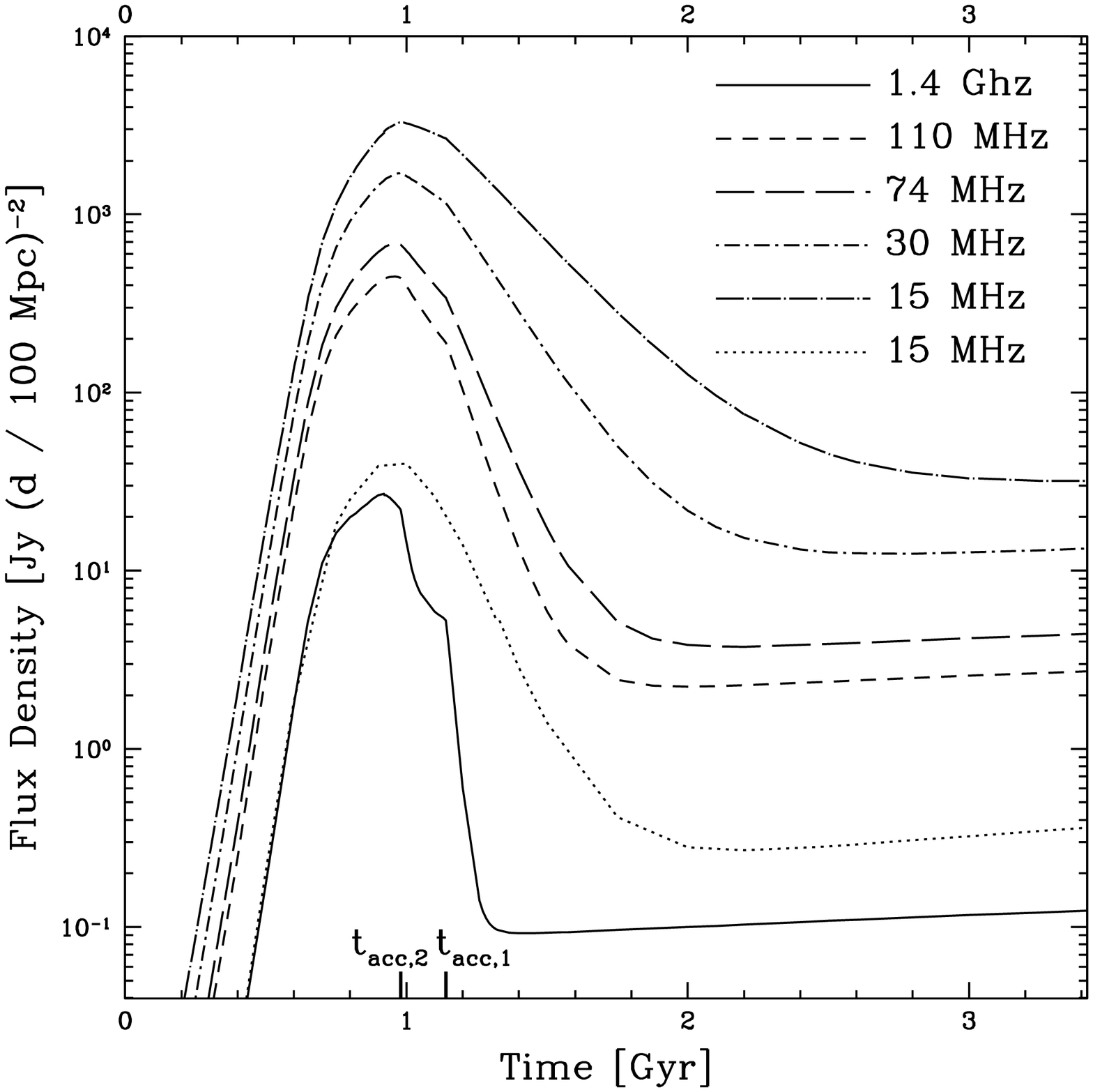}
\caption{\small Light curves at various observing frequencies produced by a
shock formed in a merger between $10^{14}$ M$_\odot$ and $10^{15}\msun$
clusters that begins at $z_i = 0.3$ and is evolved to the present epoch ($t =
3.42$ Gyr). All light curves are for a magnetic field strength of $B=1.0\mug$
unless otherwise noted.  Radio light curves in Jansky units are given at 15
MHz, 30 MHz, 74 MHz, 110 MHz, and 1.4 GHz.  The 15 MHz light curve is also
calculated with a magnetic field strength of $B=0.1\mug$ ({\em dotted
curve}).}
\label{fig:light_curves_radio}
\end{center}
\end{figure}

\section{Results}

\begin{figure}
\begin{center}
\includegraphics[width=\columnwidth]{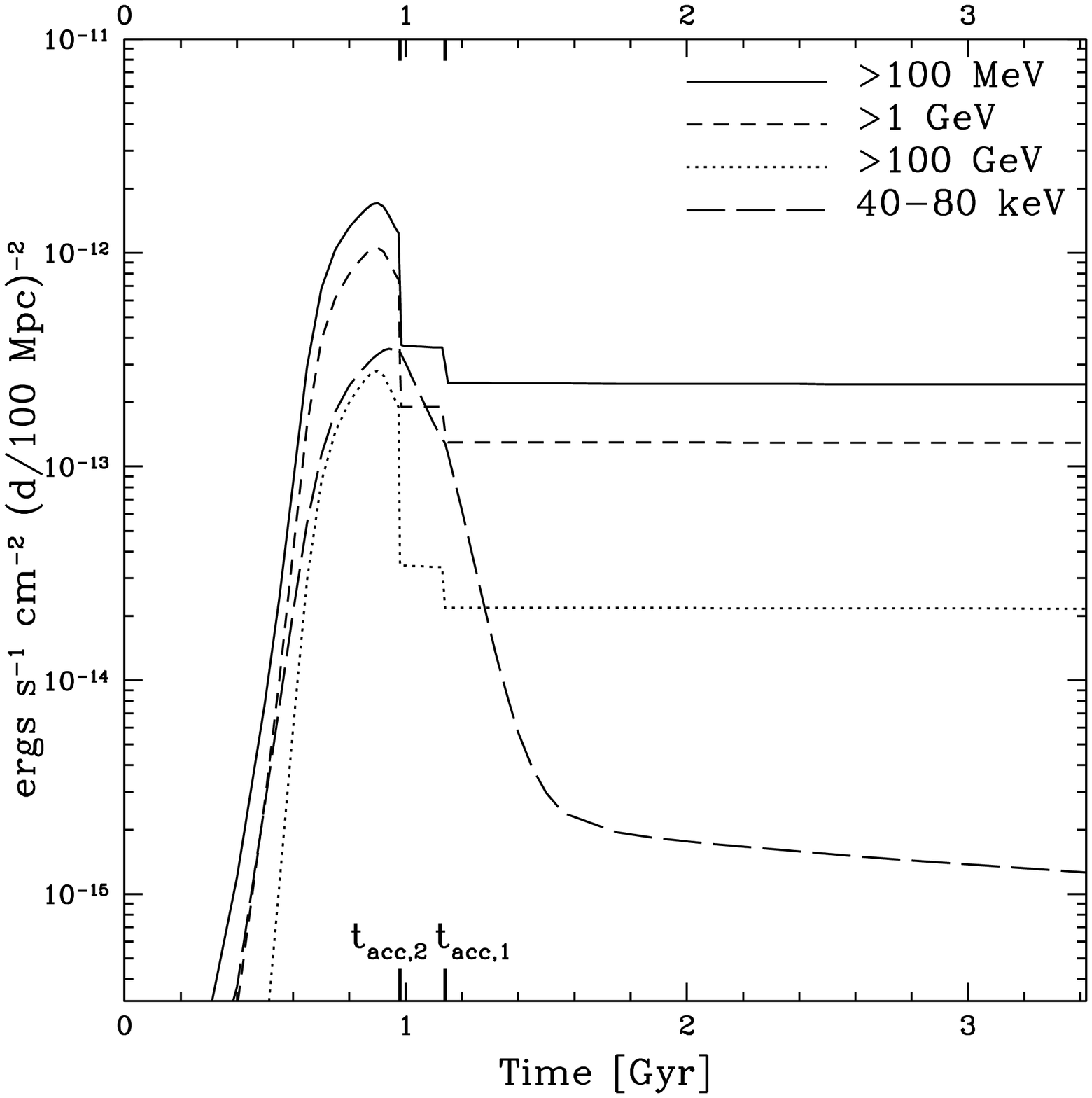}
\caption{\small Same as Figure \ref{fig:light_curves_radio} except light
curves in energy flux units are given at 40-80 keV, $>\!\!100$ MeV, $>\!\!1$
GeV and $>\!\!100$ GeV.  Curves are labeled in the diagram.}
\label{fig:light_curves_highe}
\end{center}
\end{figure}

Light curves at various observing frequencies are shown in Figures
\ref{fig:light_curves_radio} \& \ref{fig:light_curves_highe} using the
parameters for the standard case of a merger event between a dominant cluster
of mass $M_{1} = 10^{15}\msun$, and a merging cluster of mass $M_{2} =
10^{14}\msun$ with a magnetic field strength $B = 1.0\mug$, and a beginning
redshift of $z_i = 0.3$.  The light curves of the nonthermal radiation exhibit
a common behavior independent of frequency. At early times, the spectral power
rises rapidly as the clusters merge. The peak emission occurs when the centers
of mass of the two clusters pass at $t_{\rm coll}$, after which the emission
exhibits a slow decay and approaches a plateau at times $t \gtorder t_{\rm
acc}$.  The time $t_{\rm acc,1(2)}$ is defined as the time at which particle
injection has terminated at the forward shock (1) or reverse shock (2).  The
rate of decay of the emission increases with radio frequency due to the
stronger cooling of the higher energy electrons, so that the decay is slowest
at lower frequencies.  Synchrotron emission from secondary electrons forms the
late-time plateaus at radio energies.  This behavior is also apparent for the
hard X-ray emission, although it is formed by primary bremsstrahlung and both
primary and secondary Compton radiation at late times.  At $\gamma$-ray
energies, the $\pi^0$-decay emission forms a plateau of emission that
dominates soon after $t_{\rm acc}$.

\begin{figure}
\begin{center}
\includegraphics[width=\columnwidth]{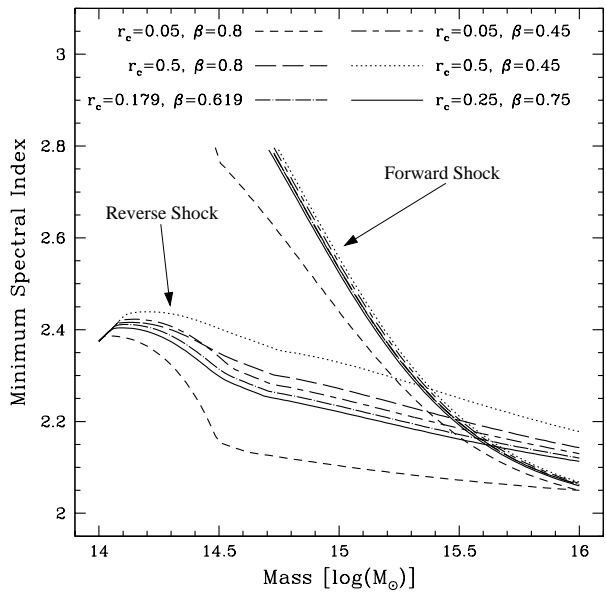}
\caption{\small Calculations of the hardest particle injection spectral
indices $s_{\rm min}$ formed in cluster merger shocks as a function of the
larger mass $M_1$ of the two clusters, for various values of $r_c$ and
$\beta$.  The values of $r_c$ are given in $\mpc$.  The minimum spectral index
for the forward and reverse shock is shown in the figure. }
\label{fig:minimum_spectral_index}
\end{center}
\end{figure}

Calculations of the hardest particle injection spectral index $s_{\rm min}$
formed in cluster merger shocks are shown in Figure
\ref{fig:minimum_spectral_index} as a function of the larger mass $M_{1}$ of
the two clusters, with a constant subcluster mass $M_2 =10^{14}\msun$.  We
also assume that the onset of the merger begins at redshift $z_i = 0.1$;
softer injection indices are obtained for mergers at larger values of $z_i$
because of the smaller maximum separations for merger events occurring at
higher redshift.  The mean cluster masses are smaller at higher redshift.
These lower mass clusters have a lower virial temperature.  If this is
considered in the calculation of $s_{\rm min}$, then it is possible that
stronger shocks may be seen at higher redshifts.  We calculate $s_{\rm min}$
for various values of $r_c$ and $\beta$. The values $(r_c,\beta) = (0.05,0.8),
(0.05,0.45), (0.5, 0.8)$, and $(0.5,0.45)$ roughly correspond to the extrema
in the range of these parameters measured for 45 X-ray clusters observed with
{\it ROSAT} \citep{wu:00}. Also shown are values of $s_{\rm min}$ for
$(r_c,\beta) = (0.179, 0.619)$, which are the average values of these
parameters for the 45 X-ray clusters, and $(r_c,\beta) = (0.25, 0.75)$, which
are the standard parameters used in the calculations.  It should be noted that
as the dominant cluster approaches the mass of the merging cluster
($10^{14}\msun$), the spectral indices for the forward and reverse shock will
be identical if the matter profiles are identical.  However, in Figure
\ref{fig:minimum_spectral_index} the values of $(r_c,\beta)$ are held
constant, and therefore; the matter profiles differ.

\subsection{Modeling the Coma Cluster}

Both optical \citep{colless:96,biviano:96,edwards:02} and X-ray
\citep{vikhlimin:97,arnaud:01} observations indicate that the dynamics of the
Coma cluster is well-described by a three-body merger
model. \citet{colless:96} were the first to find substructure in the central
region of Coma that is consistent with a recent merger event near the
collision time $t_{\rm coll}$, defined to be the time when the centers of mass
of the two clusters pass through each other.  Total mass estimates of the
dominant cluster $M_{1}$ is estimated to be $0.8 \times 10^{15}\msun$.  X-ray
observations \citep{vikhlimin:97} estimate the total mass of the merging
cluster $M_{2}$ from the gas striped in the merging process to be $\sim\!\!0.1
\times 10^{15}\msun$.  This assumes a gas fraction of $5$--$10\%$.

\begin{figure}
\begin{center}
\includegraphics[width=\columnwidth]{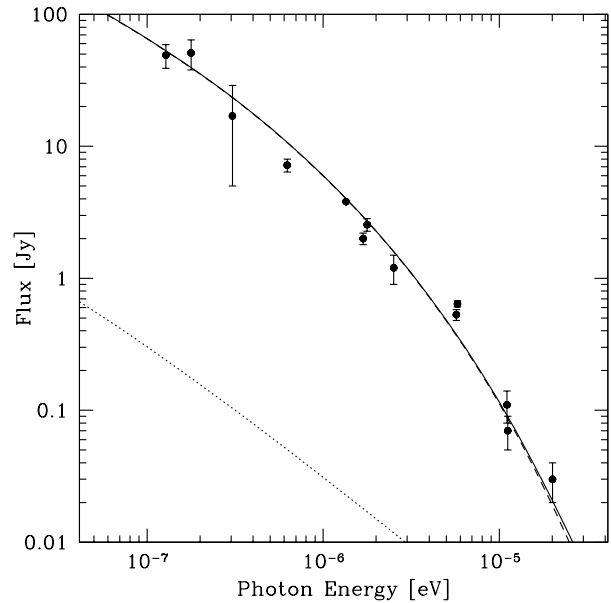}
\caption{\small Comparison of radio observations of the Coma cluster with
 cluster merger shock model.  The solid curve is the total expected radio
 emission, the dashed curve is the contribution to the total emission from
 primary electrons, and the dotted curve from secondary electrons and
 positrons.  The solid circles are observational data points taken from
 \citet{thierbach:03}. }
\label{fig:coma_radio}
\end{center}
\end{figure}

The ICM is well described by an isothermal beta model with core radius $r_{c}
= 0.257 \mpc$, central electron density $\rho_{e0} = 3.82 \times
10^{-3}\percc$, central proton density $\rho_{0} = 7.43 \times 10^{-27}
\gmpercc$, power-law slope $\beta = 0.705$, and a mean gas temperature
$\langle T_{\rm X} \rangle = 8.21\kev$ \citep{mohr:99}.  The assumed magnetic
field strength is $B=0.22\mug$, and an efficiency factor $\eta_{e,p} = 1\%$.
Because we can never know the true gas distribution in the merging cluster, we
approximate the ICM of the merging cluster by an isothermal beta model with
core radius $r_{c} = 0.150 \mpc$, central electron density $\rho_{e0} = 1.0
\times 10^{-3}\percc$, central proton density $\rho_{0} = 1.67 \times 10^{-27}
\gmpercc$, power-law slope $\beta = 0.7$, and a mean gas temperature
calculated from equation 9 in \citet{berrington:03}.

The model is evolved to a time $t=0.97\times10^{9}\yrs$, which is just prior
to the collision time $t_{\rm coll} = 1.0\times10^{9}\yrs$.  The redshift of
the cluster at the time corresponding to the creation of the shock front
corresponds to $z_{i} = 0.10$ for a $(\Omega_{0},\Omega_{\Lambda}) =
(0.3,0.7)$ cosmology.  The observed redshift of the evolved cluster is
$z=0.0232$.

\begin{figure}
\begin{center}
\includegraphics[width=\columnwidth]{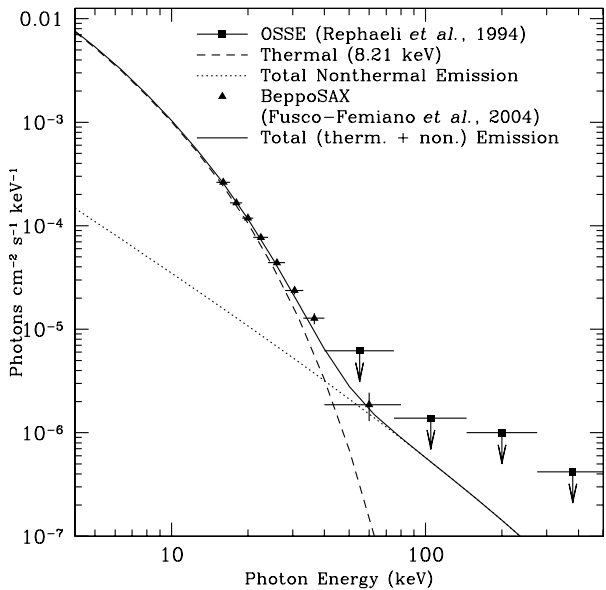}
\caption{\small Comparison of X-ray observations of the 
Coma cluster with the cluster merger shock model.  Points are labeled
in the diagram.  The thermal bremsstrahlung emission (dashed curve) is
the calculated thermal bremsstrahlung for the parameters described in
the text.  The OSSE data points are $2\sigma$ upper limits.  The solid
curve is the sum of the non-thermal (dotted curve) and the thermal
$8.21\kev$ bremsstrahlung emission.  }
\label{fig:coma_xray}
\end{center}
\end{figure}

In Figure \ref{fig:coma_radio}, we show a comparison of our model with the
observed radio emission from the radio halo \citet{thierbach:03}.  Our models
favor a primary electron source for the radio emission from the radio halo.
Despite our models using a uniform density profile for the calculation of the
secondary electron production, the emission from the secondary electron is an
upper limit to the true secondary electron emission.  However, the density in
the central region is roughly constant, so that the uniform density assumption
is a good approximation to the true secondary electron emission.

In Figure \ref{fig:coma_xray}, we show a comparison of the calculated thermal
and non-thermal emission with the observed Hard X-ray (HXR) emission from the
central region of the Coma cluster of galaxies observed by the Phoswich
Detection System (PDS) on {\it BeppoSAX} \citep{fusco-femiano:04}.  The
reported non-thermal photon emission is the total integrated emission expected
within $1.5\mpc$ of the cluster center.  This corresponds to a field of view
of $\sim1^{\circ}\!\!.7$.  The PDS has a FWHM field of view of $\sim\!\!
1^{\circ}\!\!.3$ which corresponds to a linear scale of $\sim\!\! 2.2\mpc$.
In addition we also show the OSSE $2\sigma$ upper limits \citep{rephaeli:94}.
The HXR emission observed between $20$--$80\kev$ is dominated by Compton up
scattering of photons off of primary electrons.

\begin{figure}
\begin{center}
\includegraphics[width=\columnwidth]{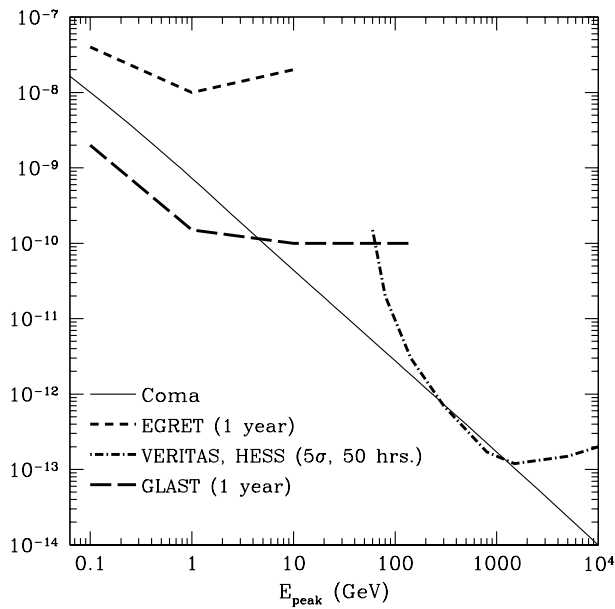}
\caption{\small Predicted $\gamma$-ray emission from the Coma cluster of
 galaxies from the cluster merger shock model.  The solid curve is the
 integrated number of particles per cm$^{2}$ per s$^{1}$ greater than E$_{\rm
 peak}$.  Observational limits for EGRET, {\it GLAST}, and VERITAS and HESS
 are included.  The EGRET and predicted {\it GLAST} observational limits are 1
 year integrations.  The quoted VERITAS and HESS limits are 50 hour, $5\sigma$
 limits \citep{weekes:02}.  The vertical axis represents the number of
 particles greater than E$_{\rm peak}$ in units of cm$^{-2}$ s$^{-1}$.}
\label{fig:coma_highe}
\end{center}
\end{figure}

The {\it BeppoSAX}/PDS observations include the merging cluster associated
with NGC 4839, and will be contaminated with any non-thermal emission
associated with shocks formed in its merging process.  The expected
non-thermal emission resulting from shocks for a cluster infalling at a
minimum distance of $1.6h^{-1}_{50}\mpc$ \citep{neumann:01} will be negligible
in comparison with the emission from the merger observed in the core of Coma
\citep{berrington:03}.

In Figure \ref{fig:coma_highe}, we present the predicted $\gamma$-ray emission
from the Coma cluster of galaxies.  The observational limits are taken from
\citet{weekes:02}.  As seen from the figure, the predicted $\gamma$-ray
emission falls comfortably below the predicted upper limits for the EGRET
observations.  Our model predicts that the space-based observatory {\it GLAST}
will detect the non-thermal $\gamma$-rays at high significance.  Furthermore,
we predict that both VERITAS and HESS will have strong $\approx 5\sigma$
detections in 50 hours of observations.

\section{Summary and Conclusions}

We describe a computer model designed to calculate the nonthermal particle
distributions and photon spectra resulting from nonthermal processes produced
by shocks that form between merging clusters of galaxies.  Over the lifetime
of a cluster merger shock $\sim\!\!10^{61}$-$10^{62}\ergs$ will be deposited
in the energy of a nonthermal particle population.  From the calculated peak
luminosities shown in Figures \ref{fig:light_curves_radio} \&
\ref{fig:light_curves_highe}, we estimate that the nonthermal emission from
merger shocks will be detected at radio frequencies with a Long Wavelength
Array out to a distance of $\sim\!\!2000\mpc$ at $15\mhz$ and
$\sim\!\!700\mpc$ at $120\mhz$.  At $\gamma$-ray energies, we estimate the
distance threshold for detection of cluster merger shocks to be
$\sim\!\!200\mpc$ for the space-based observatory {\it GLAST}.  However, each
cluster must be considered on a case by case basis to determine its
observability.

It was suggested that Galaxy clusters are a dominant contributor to the
diffuse extragalactic $\gamma$-ray background (DEGB) \citep{loeb:00}.  Our
models do not support this claim.  Shocks formed in the merger process are
weak resulting in a spectral index that is softer than the observed
$2.30\pm0.03$ index for the DEGB.  The superposition of softer spectra on the
harder spectra observed for the most massive merger events will result in a
concave spectrum at lower energies that is not observed.  We estimate that the
contribution to the DEGB by cluster merger shocks will not exceed
$\sim\!\!1$--$10\%$.  

We applied this model to the Coma cluster of galaxies.  We show that the radio
emission from the radio halo Coma C is well fit by the cluster merger model.
The calculated nonthermal X-ray emission also fits the observed HXR emission
observed at 40--80$\kev$ by {\it BeppoSAX}/PDS.  The $\gamma$-ray emission
expected from this model is also calculated and is shown to fall below the
observational limits for EGRET as reported by \citet{rei03}, but should be
strongly detected by the space-based observatory {\it GLAST} and the
ground-based air Cherenkov telescopes such as VERITAS and HESS.  Even though
other acceleration mechanisms or point sources could produce nonthermal
emission in the core of Coma, our model of cluster merger shocks account for
the entire observed radio and HXR emission.

\section{Acknowledgments}
The work of CD is supported by the Office of Naval Research and {\it
GLAST} Science Investigation Grant No.\ DPR-S-1563-Y.

\end{document}